\documentstyle[epsfig,prl,twocolumn,aps,floats]{revtex} 
\begin{document}
\topmargin = -2.0cm
\overfullrule 0pt
\twocolumn[\hsize\textwidth\columnwidth\hsize\csname
@twocolumnfalse\endcsname
\title{On the Size of the Dark Side of the Solar Neutrino Parameter
Space}
%\vglue -0.5cm
%\hfill{\vbox{ \hbox{hep-ph/0002186}  
%\hbox{IFIC/00-16}}}
\vglue 0.5cm
\author{M.\ C.\ Gonzalez-Garcia and C. Pe\~na-Garay}
\address{\sl  Instituto de F\'{\i}sica Corpuscular -- C.S.I.C. \\
 Departamento de F\'{\i}sica Te\`orica, Universitat de Val\`encia \\
Edificio Institutos de Paterna, Apt 2085, 46071 València, Spain }
\maketitle
\vspace{.5cm}
\hfuzz=25pt
\begin{abstract} 
We present an analysis of the MSW neutrino oscillation  
solutions of the solar neutrino problem in the framework of  
two--neutrino  mixing in the enlarged parameter space   
$(\Delta m^2, \tan^2\theta)$  with $\theta \in (0,\frac{\pi}{2})$.
Recently, it was pointed out that the allowed region of parameters  
from a fit to the measured total rates can extend  
to values $\theta \geq \frac{\pi}{4}$ (the so called ``dark side'') 
when higher confidence levels are allowed.
The purpose of this letter is to reanalize the problem 
including all the solar neutrino data available, to discuss 
the dependence on the statistical criteria in the determination of the
CL of the ``dark side'' and to extract the corresponding limits 
on the largest mixing allowed by the data. Our results show that when
the Super--Kamiokande data on the zenith angle distribution of events 
and the spectrum information is included, the regions extend more into 
the second octant. 
\end{abstract}
\vskip2pc]
\newpage

\medskip

In a recent work, de Gouvea et al.~\cite{deGouvea} have pointed out that the
study of two--active neutrino oscillations in the framework of the 
MSW \cite{msw} solutions to the solar neutrino deficit as done 
traditionally on the 
$(\Delta m^2,\sin^2(2\theta)$ parameter space 
is incomplete since it covers only the range
$0\leq \theta\leq\frac{\pi}{4}$ (which they denote as ``light side''). 
By fitting the data on the total rates meassured at Chlorine \cite{homestake},
 Gallium \cite{gallex,sage} and Super--Kamiokande \cite{sk99} experiments,
they claim that the allowed region of parameters extends 
to the region with $\theta\geq\frac{\pi}{4}$  (denoted as
the ``dark side'') at some reasonable confidence level (CL). 
In fact, to our knowledge, the need of extending the mixing parameter
space was first discussed in Ref.~\cite{three} where 
the mixing variable $\tan^2\theta$ was introduced to chart the full 
mixing range $0\leq\theta\leq\frac{\pi}{2}$.  

In this note we revisit the problem of the extension of the ``dark side'', 
that we will denote simply as second octant, of the solar neutrino 
parameter space after including the effect of all other solar neutrino 
observables. In particular in this analysis we use all measured total 
event rates as 
well as the 825-day Super--Kamiokande data on the zenith angle dependence and 
the recoil electron energy spectrum of the events. We also discuss the
dependence on the statistical criteria used in the construction of the 
allowed regions. Our results are summarized in Tables~\ref{lma} and~\ref{low}
where we show the maximum allowed values of $\tan^2\theta$ at the 
90 and 99\% CL when the different observables are included as well 
as the CL for the second octant of the parameter space. We show that when
adding the zenith angle and the spectrum information in the analysis
the regions extend more into the second octant. In particular for the 
LMA solution, this behaviour is mainly driven by the data on 
the zenith angular dependence since the best fit point for the zenith angle 
distribution is in the second octant as pointed out in Ref.~\cite{ourfour}.
For details on the data and statistical analysis employed in this study 
we refer to our detailed work on Ref.\cite{ourtwo} and references therein.

Let's first recall that in the framework of two massive neutrinos, the
weak eigenstates ($\nu_e$ and $\nu_x$ for the solar neutrino problem) 
can be writen as a linear combination of the mass eigenstates 
$\nu_1$ and $\nu_2$ 
\begin{eqnarray}
\nu_e &=& \cos \theta ~\nu_1 
+ \sin \theta ~\nu_2 \;  \nonumber \\
\nu_x 
&= &- \sin \theta ~\nu_1 + \cos \theta ~\nu_2 ~,
\label{eigendef}
\end{eqnarray} 
where $\theta$ is refered to as  the mixing angle in vacuum. 
The mass--squared difference is defined as $\Delta m^2=m_2^2-m_1^2$.
For the solar neutrino problem $\nu_x$ can label either an active 
neutrino $x=\mu,\tau$ or an sterile neutrino. In what follows we
restrict our discussion to oscillations into active neutrinos.
For oscillations into sterile neutrinos large mixing solutions are not 
allowed~\cite{ourtwo}. 

By inspection of the symmetry properties of Eq.~(\ref{eigendef}) 
one sees that the full parameter space can be exhausted by using 
the mass-squared difference $\Delta{m}^2$ always positive and 
the mixing angle in the interval $0 \leq \theta \leq \frac{\pi}{2}$.
In the case of vacuum oscillations, moreover, the transition probabilities 
as directly derived from Eq.~(\ref{eigendef}) can be writen in terms of 
$\sin^2(2\theta)$ and therefore they are symmetric under the change
$\theta \rightarrow \theta-\frac{\pi}{2}$ 
so each allowed value of $\sin^2(2\theta)$
corresponds to two allowed values of $\theta$.

On the other hand, in the case of the MSW solutions
the transition probability takes the form 
\begin{equation}
P_{ee} = P_{e1}^{Sun} P_{1e}^{Earth} + P_{e2}^{Sun} P_{2e}^{Earth}
\end{equation}
where $ P_{e1}^{Sun} $ is the probability that a solar neutrino, that
is created as $\nu_e$, leaves the Sun as a mass eigenstate $\nu_1$,
and $ P_{1e}^{Earth} $ is the probability that a neutrino which enters
the Earth as $\nu_1$ arrives at the detector as $\nu_e$ \cite{earth}. Similar
definitions apply to $P_{e2}^{Sun}$ and $P_{2e}^{Earth}$.
For $P_{ie}^{Earth}$ we integrate numerically the evolution equation 
in matter using the Earth density profile given in the Preliminary 
Reference Earth Model (PREM) \cite{PREM}.

The quantity $P_{e1}^{Sun}$ is given, after discarding the oscillation
terms, as
\begin{equation}
P_{e1}^{Sun}  = 1- P_{e2}^{Sun}  = 
\frac{1}{2} + (\frac{1}{2} - P_{LZ})\cos(2\theta_{m,0})
\label{psun}
\end{equation}
where $P_{LZ}$ denotes the standard Landau-Zener probability
\cite{LZ} and $\theta_{m,0}$ is the mixing angle in matter
at the neutrino production point:
\begin{eqnarray} 
\cos(2\theta_{m,0})& = & 
\frac {\Delta{m}^2 \cos(2\theta) - A_0} 
{\sqrt{ (\Delta{m}^2 \cos(2\theta) - A_0)^2 
+(\Delta{m}^2 \sin(2\theta))^2 }} %\label{thetam} 
\nonumber 
\\
P_{LZ}&=&\frac{\exp[-\gamma \sin^2\theta]-\exp[-\gamma]}{1-\exp [- \gamma]} 
\label{PLZ}   
\\
\gamma &= & \pi \frac{\Delta{m}^2}{E} 
\left(
\left.\frac{d\ln N_e(r)}{dr}\right|_{r=r_{res}}\right)^{-1}
\nonumber
\end{eqnarray}
with $A_0= 2 \sqrt{2} G_F E N_e(r_0)$ where $N_e(r_0)$ is
the number density of electrons in the production point, 
$E$ is the neutrino energy,$G_F$ is the Fermi constant and 
$r_{res}$ is the resonant point 
($\Delta{m}^2 \cos 2\theta=A_{res}$).
This probability is clearly not invariant under the change
$\theta \rightarrow \frac{\pi}{2}-\theta$
as resonant transitions are only possible for
values of $\theta$ smaller than $\frac{\pi}{4}$ as seen in 
Eq.~(\ref{PLZ}) what considerably supresses the transitions for 
$\theta > \frac{\pi}{4}$.
For this reason most of the earlier papers on the MSW effect that considered 
the two-neutrino mixing case used  ($\Delta m^2$ ,\, $\sin^2 2\theta $)  
as parameters in the 
fitting procedure~\cite{clasicos} with $\theta \in [0,\frac{\pi}{4}]$. 
In some cases the parameter space was represented  as 
$\frac{\displaystyle\sin^2 2\theta}{\displaystyle\cos 2\theta}$
~\cite{parke,parke1} but still assuming $\theta$ to be in the first octant.

Due to the fact that this choice does not exhaust the full   
parameter space once matter effects are included, other representations 
have been used to show the enlarged space particularly in the framework 
of three--neutrino \cite{three,threef} and four--neutrino oscillations 
\cite{ourfour}. In this way 
two suggestions to parametrize the mixing angle have been made in the 
literature:  $\sin^2\theta$ and $\tan^2 \theta$ 
with $\theta \in [0,\frac{\pi}{2}]$. For the sake of comparison with the
analysis in Ref.~\cite{deGouvea} we choose to show our results 
in terms of $\tan^2\theta$.

\begin{figure}
\begin{center}
\mbox{\epsfig{file=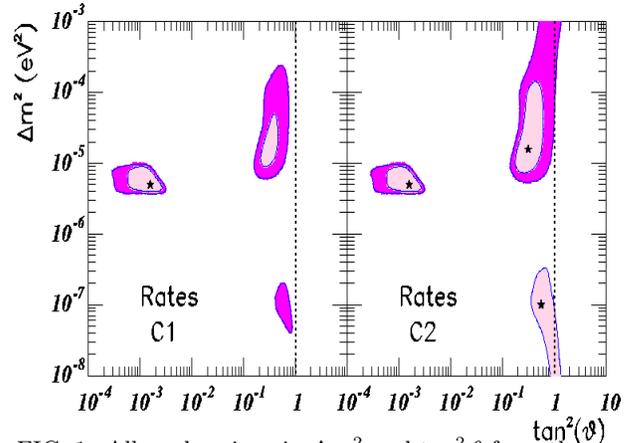,width=0.45\textwidth,height=0.25\textheight}}
\end{center}
\vglue -.5cm 
\caption{Allowed regions in  $\Delta m^2$ and $\tan^2\theta$ 
from the measurements of the total event rates at Chlorine, Gallium
and Super--Kamiokande (825-day data sample) in the two different
statistical criteria. The darker (lighter) areas indicate the 99\% (90\%)CL 
regions. The best--fit point used to defined the regions are 
indicated by a star.}
\label{rates} 
\end{figure}

\begin{figure}
\begin{center}
\mbox{\epsfig{file=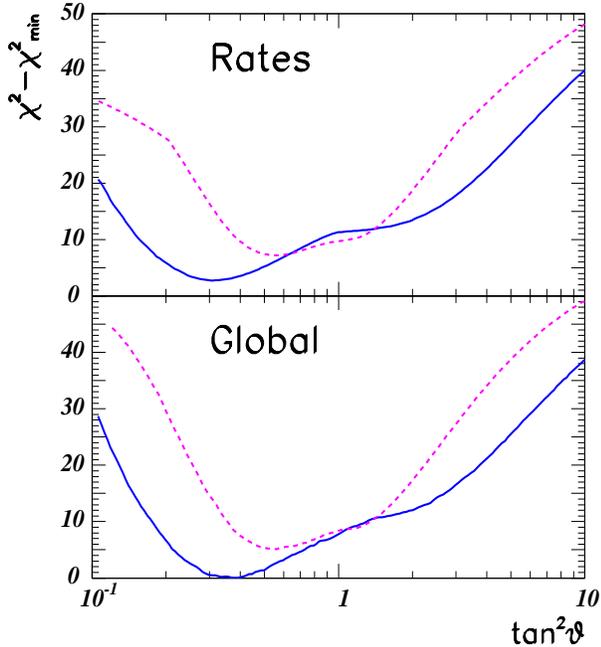,width=0.47\textwidth,height=0.4\textheight}}
\end{center}
\vglue -.2cm 
\caption{$\chi^2_{min}(\tan^2\theta)-\chi^2_{min}$ 
in the LMA (full lines) and LOW (dashed lines) regions as a fuction 
of $\tan^2\theta$,  coming from 
the fit to the total rates (upper curves) and from the global data 
analysis (lower curves). }
\label{chi2} 
\end{figure}

We now turn to our results.
We first determine the allowed range of oscillation parameters using
only the total event rates of the Chlorine, Gallium and
Super--Kamiokande experiments. We have not included
in our analysis the Kamiokande data \cite{kamioka} as it is well in
agreement with the results from the Super--Kamiokande experiment and
the precision of this last one is much higher. 
For the Gallium experiments we have used the weighted average of the results
from GALLEX \cite{gallex} and SAGE \cite{sage} detectors. Our results
are shown in Fig.~\ref{rates} where we show the allowed regions.
We choose to show our allowed regions for the ``conventional'' 
90 and 99 \% CL. It is obvious that increasing the allowed CL would lead to
larger regions. The CL for the second octant and the maximum values of 
$\tan^2 \theta$  for which the LMA and LOW solutions are allowed at 
those CL can be found in Tables~\ref{lma} and~\ref{low}. 
Furthermore, in Fig.~\ref{chi2} we show the value of 
$\chi^2_{min}(\tan^2\theta)-\chi^2_{min}$ 
in the LMA and LOW regions 
(where $\chi^2_{min}(\tan^2\theta$) is minimized in $\Delta m^2$) 
as a fuction of $\tan^2\theta$. From the figure it is possible 
to extract the maximum allowed values of $\tan^2\theta$ 
at any other CL. For instance, taking the global analysis in the LMA region 
from the figure we see that $\tan^2\theta=3$ ($\theta=60^\circ$) is 
possible only for $\Delta\chi^2 >16.6$  which corresponds 
to 99.98 for 2~dof (~3.65$\sigma$~). 
We present our results according to two different statistical criteria 
used in the literature in the defintion  of the allowed paramers.
The use of each criterion depends on the physics scenario to which the  
result of the analysis is to be applied.  
\begin{itemize} 
\item Criterion 1 (C1): The regions at certain CL are defined in terms 
of shifts of the $\chi^2$ function for 2-d.o.f, $\Delta\chi^2$=4.6 (9.2) at
90 (99) \% CL, with respect to the {\sl global minimum in the full plane}.
This criterion is used, for instance, in 
Refs.~\cite{deGouvea,sk99,bks98,ourtwo,ourfour,hata}. 
It is applicable when no region of the parameter space SMA, LMA, 
or LOW is a priory assumed to be the right one. 
\item Criterion 2 (C2): The regions at certain CL are defined in terms 
of shifts of the $\chi^2$ funcion for 2-d.o.f 
with respect to the {\sl local minimum in the corresponding region}.
This criterion holds when assuming that a given solution,  
SMA, LMA, or LOW, is the valid one. 
It clearly yields less restrictive limits.  
This criterion is used, for instance, in Refs.~\cite{bks99,ourfour}. 
\end{itemize} 
In the figures we mark with a star the location of the minima used to define 
the contours in each case. For instance, in Fig.~\ref{rates} when using 
criterion 1, the contours are defined with respect to minimum 
$\chi^2_{min}=0.37/1$~dof which is obtained in the small mixing angle 
solution. For criterion 2 we define each region with respect to its  
local minimum whose value and the corresponding goodness of the fit 
is given in Tables~\ref{lma} and~\ref{low} respectively. Notice that 
the relatively bad goodness of the fit for the LMA and LOW regions from the 
analysis of the rates only is the reason why for example in 
Fig.~\ref{rates} the LMA and LOW regions are quite different using C1 and C2.
\begin{table}
\begin{tabular}{|c|c|c|c|c|}
\hline
LMA & $\chi^2_{min,LMA}$/dof &~Crit.&$\tan^2(\theta)_{max}$ & CL (\%)\\
    &    $gof$       &       &                    &  $2^{nd}$ Octant \\
\hline
~rates&   2.9/1  &   C1          &$0.47$ ($0.76$) & $99.6$ \\
      &   91 \% 
      & C2          &$0.62$ ($1.35$)  & $98.7$ \\
\hline
~rates   & 7.0/6    & C1      &$0.62$ ($1.01$) & $99.0$ \\
+ zenith & 68 \%      & C2    &$0.69$ ($1.15$) & $98.2$ \\
\hline
~rates   & 22.1/18     &C1     &$0.55$ ($0.92$) & $99.1$ \\
+ spectrum& 77 \%      &C2     &$0.55$ ($0.92$) & $99.1$ \\
\hline
~global &  27.0/23     &C1     &$0.69$ ($1.18$) & $98.2$ \\
        &  74 \% &C2     &$0.69$ ($1.18$) & $98.2$ \\
\hline
\end{tabular}
\caption{$\chi^2_{min}$ and its gof(goodness of the fit), maximum allowed values of $\tan^2(\theta)$ at 90 (99) \% CL and the confidence level at which  
$\theta\ge \frac{\pi}{2}$ is allowed 
in the LMA region for the different set of observables and the two statistical
criteria discussed in the text.}
\label{lma}
\end{table}

As seen in Fig.~\ref{rates} and Tables~\ref{lma} and~\ref{low} when using 
criterion 1, we do not find any solution in the second octant at 99 \% CL 
from the analysis of the rates only.
One must increase the CL to 99.6 (99.6) for the LMA (LOW) region to extend 
into values $\tan^2\theta>1$. This is in agreement, for instance, with 
the results of Ref.~\cite{threef}. In Ref.~\cite{deGouvea}
they find some small allowed region in the second octant for the LOW solution
at the 99 \% CL. We have traced the origin of this small discrepancy to their
use of the exponential approximation for the electron number density 
in the sun. In our calculation we use the solar neutrino fluxes from 
Ref.\cite{BP98} and the new numerical parametrization of
the sun density as given by Bahcall~\cite{BP00}. We have explicitely
verified that if using the exponential approximation in our calculation 
the allowed LOW region extends into the second octant at 99 \% CL. 

The situation is changed when the regions are defined 
according to criterion 2, as seen in Fig.~\ref{rates} and Tables~\ref{lma} 
and~\ref{low}. In particular both the LMA and LOW regions overlap at 
99\% CL because the value of $\chi^2$ in between the two regions is below
the 99 \% CL defined respect to the local LOW miminum 
(for this reason the LOW region in the figure is only shown 
at the 90 \% CL). 
Notice also that in our representation we have chosen to cut the parameter 
space at $\Delta m^2 >10^{-8}$ eV$^2$. Recently, in Ref.~\cite{Friedland} 
it has been pointed out that matter effects may be relevant for lower 
values of $\Delta m^2$. One must notice, however that for such lower 
mass values the simple analytic expresions in Eqs.(\ref{psun}) and~(\ref{PLZ})
start loosing validity \cite{parke1,threef,krastev}. 
\begin{table}
\begin{tabular}{|c|c|c|c|c|}
\hline
LOW & $\chi^2_{min,LOW}/dof$ &~Crit.&$\tan^2(\theta)_{max}$ & CL (\%) \\
    &    $gof$       &       &                    &  $2^{nd}$ Octant \\
\hline
~rates &  7.4/1  &C1 &~-~ ($0.87$)    & $99.6$ \\
       &  99 \%   &C2 &$1.37$ ($1.78$) & $82.6$ \\
\hline
~rates&   12.7/6    &C1       &~-~ ($0.94$) & $99.1$ \\
+ zenith & 95 \% 
   &C2       &$1.35$ ($1.77$) & $75.5$ \\
\hline
~rates&   26.7/18      &C1  &~-~ ($1.29$) & $98.1$ \\
+ spectrum & 92 \%   &C2  &$1.23$ ($1.69$) & $84.3$ \\
\hline
~global&  32./23  & C1   &~-~ ($1.24$) & $98.3$ \\
       &  90  \%  & C2    &$1.30$ ($1.74$) & $79.6$ \\
\hline
\end{tabular}
\caption{$\chi^2_{min}$ and its gof(goodness of the fit), maximum allowed values of $\tan^2(\theta)$ at 90 (99) \% CL and the confidence level at which $\theta\ge \frac{\pi}{2}$ is allowed in the LOW region for the different set of observables and the two statistical criteria discussed in the text.}
\label{low}
\end{table}

\begin{figure}
\begin{center}
\mbox{\epsfig{file=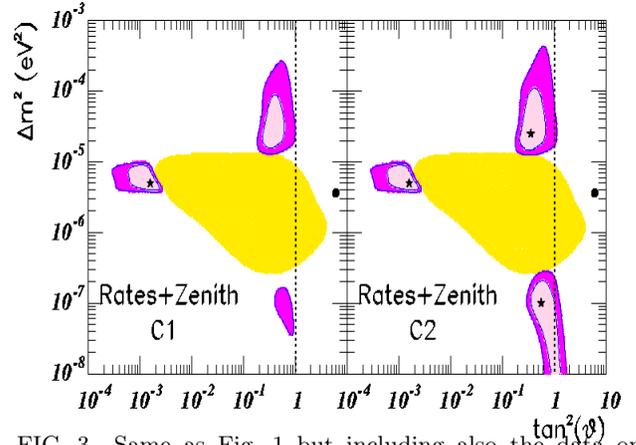,width=0.45\textwidth,height=0.25\textheight}}
\end{center}
\vglue -.5cm 
\caption{Same as Fig.~\protect\ref{rates} 
but including also the data on the zenith
angle distribution observed in Super--Kamiokande. 
The shadowed area represents the
region excluded by the zenith angle distribution data at 99\% CL. 
The location of minimum of the $\chi^2$ function for the zenith angle data is
market with a point.}
\label{ratesz} 
\end{figure}

\begin{figure}
\begin{center} 3 
\mbox{\epsfig{file=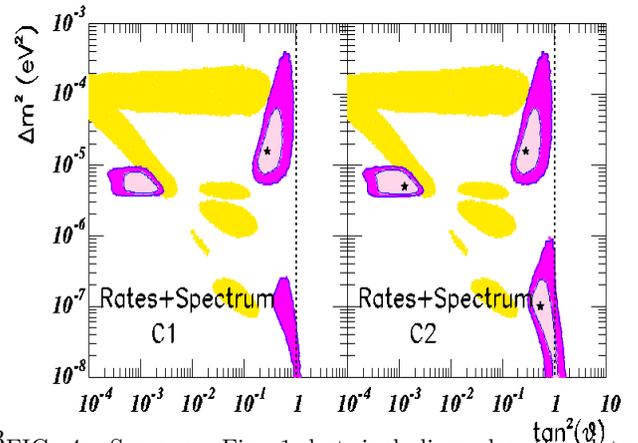,width=0.45\textwidth,height=0.25\textheight}}
\end{center}
\vglue -.5cm 
\caption{Same as Fig.~\protect\ref{rates} but including also the data on the recoil electron energy spectrum 
observed in Super--Kamiokande. 
The shadowed area represents the region excluded by spectrum  data at 99\% CL.
} 
\label{ratess}
\end{figure}

In Figs.~\ref{ratesz} and~\ref{ratess}
we show the allowed regions when either
the data on the zenith angular dependence or the recoil electron energy 
spectrum  are combined with the results from the total rates. 
The corresponding values of the absolute minimum of the  $\chi^2$ 
fuction for the combination of rates plus zenith angular dependence
data (rates plus recoil electron energy spectrum)   
are $\chi^2_{min}=5.9/3$~dof (22.1/15~dof) which are obtained 
for the SMA (LMA) solutions
and are used in the construction of the allowed regions for criterion 1.
In the figures we also show the corresponding excluded regions at 99 \%CL 
by the new observables. 

As seen in Fig.~\ref{ratesz} and in the tables 
the inclusion of the data on the zenith angular dependence of the
Super--Kamiokande events leads, in general, to  
an increase in the maximum allowed values of $\tan^2\theta$ and a better 
CL for the second octant for both 
LMA and LOW regions.
This is due to the fact that the best fit point for the angular distribution
is obtained in the second octant ($\Delta{m}^2_{21}=
3.7\times 10^{-6}$ eV$^2$, $\tan^2(\theta)=5.9$ with 
$\chi^2_{min,zen}=1.5$) as shown in Fig.~\ref{ratesz} \cite{ourfour}.

This fact has been obviated in past analysis in the two neutrino 
oscillations and it must be taken into account to do properly the 
$\chi^2$ analysis when the data on zenith dependence is included.
In this respect, one must notice, for instance, that in their
preliminary analysis on the zenith angle dependence \cite{sk99} 
SuperKamiokande obtains the minimum $\chi^2$ at the boundary of their 
``cut'' mixing parameter space $\sin^2(2\theta)=1$. In this way, their 
$\chi^2_{min}$ is higher than the ``true'' minimum which is missed. 
Although, at present, the difference in the excluded region 
defined with respect the ``true'' or with respect to  the ``cut'' minimum
is small, this can be of further importance when more data is acummulated.

We have also explicitely verified that the results on the zenith angle 
excluded region as well as the best fit point
position are very mildly dependent on the exact profile of the Earth.
Very similar results can be obtained by using the analytical expressions
valid for the two--step Earth density profile \cite{earth}.

\begin{figure}
\begin{center}
\mbox{\epsfig{file=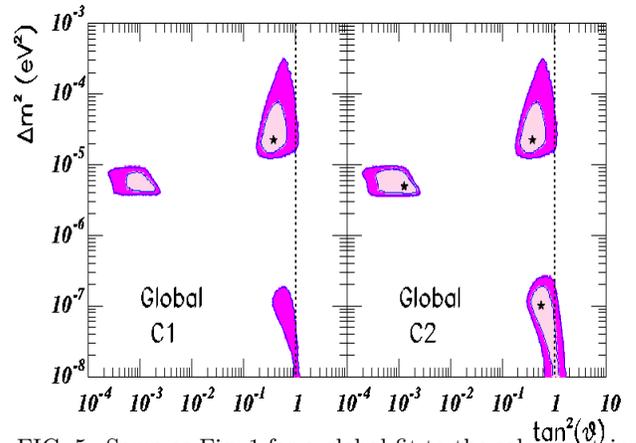,width=0.45\textwidth,height=0.25\textheight}}
\end{center}
\vglue -.5cm 
\caption{Same as Fig.~\protect\ref{rates} for a global fit to the 
solar neutrino data.}
\label{global}
\end{figure}

The effect of the inclusion of the recoil electron energy 
spectrum is shown in Fig.~\ref{ratess}. In this case when using criterion 1  
we also find  
an increase in the maximum allowed values of $\tan^2\theta$  and a better 
CL for the second octant for both  LMA and LOW regions as compared to the results for the analysis of the rates only. However this is not the case when using
criterion 2 as can be seen in the tables. 
The allowed regions from the global analysis are displayed in 
Fig.~\ref{global}. We see from the figure that at 99 \% CL both the
LMA and the LOW regions extend into the second octant when using any of the
two statistical criteria. 
In Fig.~\ref{chi2} we show the value of 
$\chi^2_{min}(\tan^2\theta)-\chi^2_{min}$ 
in the LMA and LOW  regions 
(where $\chi^2_{min}(\tan^2\theta$) is minimized in $\Delta m^2$) 
as a fuction of $\tan^2\theta$ coming from the fit to the total rates 
(for which  we recall that $\chi^2_{min}=0.37$) and to the global data set
(for which $\chi^2_{min}=\chi^2_{min,LMA}=27.0$).
As seen in the figure the inclusion of the new observables
leads to a small shift in the position of the local best fit points
towards slightly larger values of $\tan^2\theta$ . 
However, as discussed before this not always
translates into an increase on the maximum allowed value of $\tan^2\theta$ 
as the minimum also becomes deeper.

To summarize, in this paper we have studied the extension of the second octant 
of the solar neutrino parameter space after including in the analysis 
all measured total event rates as well as all 
the 825-day Super--Kamiokande data on the zenith angle dependence and 
the recoil electron energy spectrum of the events. 
We also have discuss the dependence of the results on the statistical 
criterion used in the construction of the allowed regions. Our results are 
summarized in Tables~\ref{lma} and~\ref{low}
where we show the maximum allowed values of the $\tan\theta$ at 
90 and 99\% CL when the different observables are included as well 
as the CL for the second octant of the parameter space. We have shown that when
the zenith angle and the spectrum information is included 
the regions extend slightly more into the second octant. For the LMA 
this behaviour is mainly driven by the data on the zenith
angular dependence since the best fit point for the zenith angle distribution 
is in the second octant. 

Finally just to comment that, the existence of solutions to the 
solar neutrino problem for $\theta>\frac{\pi}{4}$ is not only of
academic interest to the extent that in general they are  
perfectly allowed by the models of neutrino masses \cite{models}. 
A particular interesting recent example of a predictive model can be found 
in Ref.~\cite{rparity} where, despite the predictivity of the model, 
solutions in both octants are equally possible. Furthermore, in the 
context of three neutrino mixing, the determination of the sign 
of $\Delta m^2$ of the LMA solution can be of relevance in the exact 
determination of the $CP$-violating phase \cite{nufact}.

\acknowledgments 
We thank Carlo Giunti, H. Murayama and J. W. F. Valle for discussions. 
We are specially endebted to J. Bahcall for providing us with the
latest parametrization of the sun matter density. 
This work was supported by DGICYT under grants PB98-0693 and PB97-1261, 
and by the TMR network grant ERBFMRXCT960090 of the European Union.

\end{document}